\documentstyle[stwol]{article}

\def\Journal#1#2#3#4{{#1} {\bf #2}, #3 (#4)}

\def\NPB{{\em Nucl. Phys.} B}
\def\PLB{{\em Phys. Lett.}  B}

\def\PRD{{\em Phys. Rev.} D}
\def\ZPC{{\em Z. Phys.} C}

\def\be{\begin{equation}}
\def\ee{\end{equation}}
\def\bea{\begin{eqnarray}}
\def\eea{\end{eqnarray}}

\begin{document}

\title{Hadronic Decays and Lifetimes of B and D Mesons}

\author{T.E. Browder }

\address{University of Hawaii at Manoa,
Honolulu, HI 96822, U.S.A.}

\author {University of Hawaii Preprint UH-511-857-96}

\author{(To appear in the Proceedings of the 1996 Warsaw ICHEP Conference)}


\twocolumn[\maketitle\abstracts{We review recent experimental
results on hadronic decays and lifetimes of hadrons containing $b$
and $c$ quarks.\cite{bhp} We discuss charm counting and the
semileptonic branching fraction in B decays, 
the color suppressed amplitude in B and D decay, 
and the search for gluonic penguins in B decay.}]

\section{Charm counting and the semileptonic branching fraction}
\subsection{The Experimental Observations}\label{subsec:prod}

A complete picture of inclusive
$B$ decay is beginning to emerge from recent measurements 
by CLEO~II and the LEP experiments.
These measurements can be used to address
the question of whether the hadronic decay of the B meson is compatible with
its semileptonic branching fraction.

Three facts emerge from the experimental examination of inclusive $B$ decay:
$$n_c = 1.15 \pm 0.05$$ where $n_c$ is the number of charm quarks
produced per $B$ decay taking an average of ARGUS, CLEO 1.5, and CLEO II
results and using  ${\cal B}(D^0\to K^-\pi^+)=(3.76\pm 0.15\%)$.
$${\cal B}(B\to X\ell\nu)=10.23\pm 0.39\% $$ This value is the 
average of the CLEO and ARGUS model independent measurements using dileptons.
The third quantity is calculated from the inclusive $B\to D_s$, 
$B\to (c\bar{c}) X$, and $B\to \Xi_c$ branching fractions,
$${\cal B}(b\to c \bar{c} s)=0.158\pm 0.028\% .$$ It is determined assuming
no contribution from $D$ production, an assumption 
which can be checked using data.

\subsection{Theoretical Interpretation}\label{subsec:wpp}

In the usual parton model, it is 
difficult to accomodate a low semileptonic branching fraction unless the 
hadronic width of the B meson is increased.\cite{bsl}

The explanations for the semileptonic branching fraction
which have been proposed can be formulated by expressing
the hadronic width of the $B$ meson in terms of three components:
$$\Gamma_{hadronic}(b)  = \Gamma (b\to c \bar{c} s) 
+ \Gamma  (b\to c \bar{u} d) +\Gamma (b\to s~g).$$ If the semileptonic
branching fraction is to be reduced to the observed level, then one of these
components must be enhanced.

A large number of explanations have been proposed in the last few years.
These explanations can be logically classified as follows:

\begin{enumerate}
\item An enhancement of $b\to c \bar{c} s$ due to large QCD corrections
or the breakdown of local duality.\cite{bccs1}$^{,~}$\cite{bccs2}$^{,~}$
\cite{bccs3}$^{,~}$\cite{bccs4}

\item An enhancement of $b\to c \bar{u} d$ due to non-perturbative effects.
\cite{bcud1}$^{,~}$\cite{bcud2}$^{,~}$\cite{bcud3}$^{,~}$\cite{bcud4}

\item An enhancement of $b\to s~g$ or $b\to d~g$ 
from New Physics.\cite{new1}$^{,~}$\cite{new2}$^{,~}$\cite{new3}

\item The cocktail solution: For example, 
if both the $b\to c \bar{c} s$ and the 
$b\to c\bar{u} d$ mechanisms are increased,
this could suffice to explain the inclusive observations.

\item There might also be a 
 systematic experimental problem in the determination of either $n_c$,
${\cal B}(b\to c \bar{c} s)$, or $ {\cal B}(B\to X \ell\nu)$.\cite{isisys}
\end{enumerate}

\subsection{Other experimental clues}

Inclusive charm particle-lepton correlations can be used to probe
the $B$ decay mechanism and give further insight into this problem. 
High momentum leptons are used $p_{\ell}>1.4$ GeV
to tag the flavor of the B. The angular correlation between the meson
and the lepton is then employed to select events in which the tagging lepton
and meson are from different $B$s.

For example, the sign of $\Lambda_c$-lepton
correlations distinguishes between the 
$b\to c \bar{u} d$ and the $b\to c \bar{c} s$ mechanisms. 
Similiarly, examination of $D_s$-lepton correlations shows that
most $D_s$ mesons originate from $b\to c \bar{c} s$ rather than
from $b\to c \bar{u} d$ with $s \bar{s}$ quark popping at the lower vertex.
The same experimental 
technique can also be applied to $D$-lepton correlations.

The conventional $b\to c\bar{u} d$ mechanism which was {\it 
previously assumed} to be
responsible for all $D$ production in $B$ decay will give $D-\ell^+$
correlations. 
If a significant fraction of $D$ mesons
arise from $b\to c\bar{c} s$ with light quark popping at the
upper vertex.
This new mechanism proposed by Buchalla, Dunietz,
and Yamamoto will give $D-\ell^-$ correlations.\cite{bccs2}

Preliminary results of this study have been presented by CLEO~II which finds,
$\Gamma(B\to D~X)/\Gamma(B\to \bar{D} X) = 0.107\pm 0.029\pm 0.018$.\cite{kwon}
This implies a new contribution to the $b\to c \bar{c} s$ width
$${\cal B}(B\to D X) = 0.081\pm 0.026.$$ ALEPH finds evidence for 
$B\to D^0\bar{D^0} X + D^0 D^{\mp} X$ decays with a substantial branching 
fraction of $12.8\pm 2.7\pm 2.6 \%$.\cite{alephdd} 
DELPHI reports the observation of
$B\to D^{*+} D^{*-} X$ decays with a branching fraction 
of $1.0\pm 0.2\pm 0.3\%$.\cite{delphidd} 
Since CLEO has set upper limits on the
Cabibbo suppressed exclusive decay modes
$B\to D \bar{D}$ and $B\to D^* \bar{D^*}$ in the $10^{-3}$
range,\cite{cleodd} 
this implies that the signals observed by ALEPH and DELPHI involve
the production of a pair of $D$ mesons and additional particles.
The rate observed by ALEPH is consistent with the rate of wrong sign
$D$-lepton correlation reported by CLEO.
It is possible that these channels are actually resonant modes
of the form $B\to D {D}_s^{**} $
decays, where the p-wave  $D_s^{**}$ or radially excited $D_s^{'}$
decays to $\bar{D}^{(*)} \bar{K}$.

We can now recalculate $${\cal B}(b\to c \bar{c} s) = 0.239\pm 0.038,$$
which would suggest a larger charm yield ($n_c \sim 1.24$).
This supports
hypothesis (1),  large QCD corrections in $b\to c \bar{c} s$. 
{\it BUT the charm yield $n_c$ as computed in the usual way
is  unchanged}. The $B\to D \bar{D} K$ source was properly accounted for
in the computation of $n_c$. This suggests that the experimental
situation is still problematic.
Is there an error in the normalization ${\cal B}(D^0\to K^-\pi^+)$ 
 or is there still room for enhanced ${\cal B}(b\to c u \bar{d})$ ?

We note that ALEPH has recently reported a value for 
$n_c$ in $Z\to b\bar{b}$\cite{alephnc}. They find 
$n_c^Z = 1.230\pm 0.036 \pm 0.038 \pm 0.053$. 
The rate of $D_s$ and $\Lambda_c$ production is significantly higher
than what is observed at the $\Upsilon(4S)$. It is not clear whether
the quantity being measured is the same as $n_c$ at the $\Upsilon(4S)$, which
would be the case if the spectator model holds and if the contribution
of $B_s$ and $\Lambda_b$ could be neglected. 
OPAL has reported a somewhat lower value of $n_c$.

There are other implications of these observations.
A $B$ decay mechanism with a ${\cal O}(10\%)$ branching 
fraction has been found
which was not previously included in the CLEO or LEP
Monte Carlo simulations of $B$ decay. This may have consequences
for other analyses of particle-lepton correlations. For example,
CLEO has re-examined the model independent dilepton measurement
of ${\cal B}(B\to X \ell\nu)$. Due to the lepton threshold of 0.6 GeV
and the soft spectrum of leptons, that measurement
is fortuitously unchanged.

\section{The sign of the color suppressed amplitude and lifetimes}

The sign
and magnitude of the color suppressed amplitude can be determined using
several classes of decay modes in charm and bottom mesons. The numerical
determination assumes factorization and uses form factors from various
phenemonological models.

For $D$ decay one
uses exclusive modes such as $D\to K\pi$, $D\to K\rho$ etc., 
and obtains 
$$ a_1 = 1.10\pm 0.03,~ a_2 = -0.50\pm 0.03  $$
The destructive interference observed in two body $D^+$
decays leads to the $D^+$-$D^0$ lifetime difference.

For $B$ decay, one can find the magnitude of $|a_1|$ from
the branching fractions for the decay modes
$\bar{B}^0\to D^{(*)+}\pi^-$, $\bar{B}^0\to D^{+(*)}\rho^-$.
This gives $|a_1|=1.06\pm 0.03\pm 0.06$.
One can also extract $|a_1|$
from measurements of branching fractions
$B\to D^{+,(0)} D_s^{(*)-}$. 
The magnitude $|a_2|$ can be determined from the branching
fractions for $B\to \psi K^{(*)}$. This yields $|a_2|=0.23\pm 0.01\pm 0.01$.

The value of $a_2/a_1$ can be found by comparing
$B^-$ decays where both the external and spectator diagrams
contribute to $\bar{B}^0$ decays where only the external spectator
decays contribute.
The model of Neubert et al. predicts the following ratios:
\begin{equation}
R_1 = {{\cal B}(B^- \to D^0 \pi^-) \over {\cal B}(\bar{B^0}\to D^+ \pi^-)}
                = (1 + 1.23 a_2/a_1)^2  \label{colrate1}
\end{equation}
\begin{equation}
R_2 = {{\cal B}(B^- \to D^0 \rho^-)
\over {\cal B}(\bar{B^0} \to D^+ \rho^-)}
                = (1 + 0.66 a_2 /a_1)^2  \label{colrate2}
\end{equation}
\begin{equation}
R_3 = {{\cal B}(B^- \to D^{*0} \pi^-)
         \over {\cal B}(\bar{B^0} \to D^{*+} \pi^-)}
                     =(1 + 1.29 a_2/a_1)^2  \label{colrate3}
\end{equation}
\begin{equation}
R_4 = {{\cal B}(B^- \to D^{*0} \rho^-)
          \over{\cal B}(\bar{B^0} \to D^{*+} \rho^-)}
                     \approx (1 + 0.75 a_2/a_1)^2   \label{colrate4}
\end{equation}

Using the latest branching fractions, 
we find $$ a_2/a_1 = 0.26 \pm 0.05 \pm 0.09,$$ where the second
error is due to the uncertainty ($\sim 20\%$)
in the relative production of $B^+$ and
$B^0$ mesons at the $\Upsilon(4S)$. This
is consistent with $|a_2|$/$|a_1|$ where $|a_2|$ is computed
from $B\to \psi$ modes and $|a_1|$ is computed from $\bar{B}^0\to D^{(*)}\pi,
D^{(*)}\rho$ modes.

If the constructive interference which is observed in these
$B^-$ decays is present in all $B^-$ decays, then we expect
a significant $B^-$-$B^0$ lifetime difference ($\tau_B^{-}< \tau_{B^0}$), 
of order $15-20\%$. This is only marginally consistent 
with experimental measurements of lifetimes;
the world average computed in 
our review\cite{bhp} is $$\tau_{B^-}/\tau_{B^0}= 1.00\pm 0.05 .$$ 

It is possible 
that the hadronic $B^-$ decays that have been observed so far are 
atypical. The remaining higher multiplicity $B^-$ decays could
have destructive interference or no interference. Or perhaps 
there is a mechanism which also enhances the $\bar{B}^0$
width to compensate for the increase in the $B^-$ width
and which maintains the $B^+/B^0$ lifetime ratio near unity.
Such a mechanism would be relevant to the charm counting and
semileptonic branching fraction problem.
In either case, there will be experimental consequences in the
pattern of hadronic $B$ branching fractions.

\section{The search for the gluonic penguin}

It is important to measure the size of ${\cal A}(b\to s~g)$,
the amplitude for the gluonic penguin,  in order
to interpret the CP violating asymmetries which will be observed at
future facilities. Gluonic penguin modes will also be used to search
for direct CP violation.

CLEO-II has observed a signal in the sum of $\bar{B}^0\to K^+ \pi^-$
and $\bar{B}^0\to \pi^+ \pi^-$ with a branching fraction of
$(1.8^{+0.6+0.2}_{-0.5-0.3})\times 10^{-5}$ and for the individual modes
${\cal B}(B^0\to \pi^+\pi^-)<2.0\times 10^{-5}$,
${\cal B}(B^0\to K^+\pi^-)<1.7\times 10^{-5}$. Similiar results with
consistent branching fractions have been 
reported by DELPHI\cite{delphikpi} and ALEPH\cite{alephkpi}.
CLEO-II has also observed a signal in the sum of $B^-\to K^- \omega$
and $B^-\to \pi^- \omega$.\cite{omegah} 
The combined branching fraction is
$(2.8\pm 1.1\pm 0.5) \times 10^{-5}$. 
In all of these cases, due to the paucity of
events and the difficulty of distinguishing high momentum kaons and pions,
the conclusion is that either $b\to u$ or $b\to s~g$ decays
or a combination of the two has been observed.

Another approach using quasi-inclusive decays is described in a recent
CLEO contribution.\cite{inclusive} At the $\Upsilon(4S)$, 
two body decays from $b\to s~g$ can be distinguished
from $b\to c$ decays by examination of the inclusive particle momentum 
spectrum; the $b\to s~g$ decays populate a region beyond the kinematic limit
for $b\to c$. This approach has been applied to inclusive $\eta^{'}$, $K_s$,
and $\phi$ production.

A search for  inclusive signatures of $b\to s$ gluon rather
than exclusive signatures has two possible advantages.
The inclusive rate may be calculable from first principles and is expected to 
be at least an order
of magnitude larger than the rate for any exclusive channel.
For example, the branching fraction for $b\to s q\bar{q}$ 
(where $q=u, ~d,~s$) is
${\cal O}(1\%)$\cite{Grigjanis},\cite{Desh1} and
the branching fraction for the inclusive process
$b\to s\bar{s} s$ is expected 
to be $\sim 0.23\%$ in the Standard Model\cite{Desh1}, 
while low multiplicity decay
modes such as $\bar{B}^0\to \phi K_s$ or $\bar{B}^0\to K^-\pi^+$
are expected to have branching fractions
of order $10^{-5}$.
The disadvantage of employing an inclusive method 
is the severe continuum background that must be
subtracted or suppressed.

The  decay 
$B \to \eta^{'}  X_s$, where $X_s$ denotes a meson containing
an $s$ quark, is dominated by the gluonic penguins, 
$b\to s g^*$ $g^*\to s \bar{s}$, $g^*\to u \bar{u}$ or
$g^*\to d \bar{d}$. 
The decay $B \to K_s  X$, where $X$ denotes a meson which contains
no $s$ quark, arises from a similiar gluonic penguin, 
$b\to s g^*$ $\to s \bar{d} d$.

An analogous search for $b\to s g^*$, $g^*\to s \bar{s}$
was carried out by CLEO using 
high momentum $\phi$ production\cite{phix}. In the search for
high momentum $\phi$ production, limits were obtained using 
two complementary techniques. A
purely inclusive technique with shape cuts gave a  limit 
${\cal B}(B\to X_s \phi)<2.2 \times 10^{-4}$ for $2.0<p_{\phi}<2.6$
GeV. Using the $B$ reconstruction technique, in which combinations of
the $\phi$ candidate, a kaon, and up to 4 pions were required to be
consistent with a $B$ candidate, gave a limit of 
${\cal B}(B\to X_s \phi)<1.1 \times 10^{-4}$ for $M_{X_s}<2.0$ GeV,
corresponding to $p_{\phi}>2.1$ GeV.
These results can be compared to the
Standard Model calculation of Deshpande {\it et al.}\cite{Desh2},
which predicts that the branching fraction for this process 
should lie in the range $(0.6-2.0) \times 10^{-4}$
and that $90\%$ of the $\phi$ mesons from this mechanism will lie
in the range of the experimental search. 
Ciuchini {\it et al.}\cite{ciuchini} predict a branching fraction for
${\cal B}(B\to X_s \phi)$ in the range $(1.1\pm 0.9)\times 10^{-4}$.
One sees that the sensitivity of the inclusive method is
nearly sufficient to observe a signal from Standard Model $b\to s~g$.

Using the purely inclusive technique, 
a modest excess was observed in the signal region for 
quasi two-body $B\to \eta^{'} X_s$ decays. 
A 90\% confidence level 
 upper limit of for the momentum interval $0.39<x_{\eta^{'}}<0.52$,
$${\cal B}(B\to \eta^{'} X_s) < 1.7 \times 10^{-3}$$
is obtained. Further work is in progress to improve the sensitivity
in this channel.
Examination of high momentum $K_s$ production gives
a 90\% confidence level upper limit of 
$${\cal B}(B\to K_s X) < 7.5
\times 10^{-4}$$ for $0.4 < x_{K_s} < 0.54$. 
More theoretical work is required to convert these limits into
constraints on $b\to s~g^*, g^*\to q \bar{q}$.


\section*{References}
\begin{thebibliography}{99}
\bibitem{bhp} T.E. Browder, K. Honscheid, and D. Pedrini, UH-515-848-96,
OHSTPY-HEP-E-96-006, to appear in the 1996 edition of
 Annual Reviews of Nuclear and Particle Science.

\bibitem{browhon} T.E. Browder and K. Honscheid, Progress in Nuclear
and Particle Physics, Vol. 35, ed. K. Faessler, p. 81-220 (1995).


\bibitem{bsl} I.I. Bigi, B. Blok, M. Shifman, A. Vainshtein,
\Journal{\PLB}{323}{408}{1994}.


\bibitem{bccs1} A. Falk, M. Wise, I. Dunietz, 
\Journal{\PRD}{51}{1183}{1995}; \Journal{\PLB}{73}{1075}{1995}.

\bibitem{bccs2} M. Buchalla, I. Dunietz, H. Yamamoto, 
\Journal{\PLB}{364}{188}{1995}.

\bibitem{bccs3} E. Bagan, P. Ball, V. Braun, P. Gosdzinsky, 
\Journal{\NPB}{432}{3}{1994}; \Journal{\PLB}{342}{362}{1995} and Erratum;
\Journal{\PLB}{374}{363}{1996}.

\bibitem{bccs4} W.F.Palmer and B. Stech, \Journal{\PRD}{48}{4174}{1993}.

\bibitem{bcud1} K. Honscheid, K.R. Schubert, and R. Waldi, 
\Journal{\ZPC}{63}{117}{1994}.

\bibitem{bcud2} M. Neubert, CERN-TH-96-120 , hep-ph/9605256

\bibitem{bcud3} G. Altarelli, G. Martinelli, 
S. Petrarca, and F. Rapuano, CERN-TH-96-77, hep-ph/9604202

\bibitem{bcud4}
 I.L. Grach, I.M. Narodetskii, G. Simula, and K.A. Ter-Martirosyan,
hep-ph/9603239.


\bibitem{new1} A. L. Kagan, \Journal{\PRD}{51}{6196}{1995}.

\bibitem{new2} L. Roszkowski, M. Shifman, \Journal{\PRD}{53}{404}{1996}.

\bibitem{new3} B. Grzadowski and W.S. Hou, \Journal{\PLB}{272}{383}{1992}.

\bibitem{isisys} I. Dunietz, FERMILAB-PUB-96/104-T

\bibitem{kwon} Y. Kwon (CLEO Collaboration), contribution to the
Proceedings of the 1996 Rencontres de Moriond, Editions Frontieres.

\bibitem{alephdd} ALEPH Collaboration, ICHEP96 PA05-060

\bibitem{alephnc} D. Buskulic et al. (ALEPH Collaboration), CERN PPE 96-117,
submitted to Physics Letters B.

\bibitem{delphidd} DELPHI Collaboration, ICHEP96 PA01-108, 
DELPHI 96-97 CONF 26.

\bibitem{cleodd} M. Bishai et al. (CLEO Collaboration), 
CLEO-CONF 96-10, ICHEP96, PA05-072

\bibitem{delphikpi}  W. Adam et al. (DELPHI Collaboration), CERN-PPE 96-67

\bibitem{alephkpi} D. Buskulic et al. (ALEPH Collaboration), CERN-PPE 96-104.

\bibitem{omegah} B. Barish et al. (CLEO Collaboration),
CLEO CONF 96-23, ICHEP96 PA05-095.

\bibitem{inclusive} M. Artuso et al. (CLEO Collaboration),
 CLEO CONF 96-18, ICHEP96 PA05-73.


%
\bibitem{Grigjanis} R. Grigjanis, P.J. O'Donnell, M. Sutherland,
and H. Navelet, \Journal{\PLB}{224}{209}{1989}.

\bibitem{Desh1} N. G. Deshpande, X.-G. He 
and J. Trampetic, \Journal{\PLB}{377}{161}{1996}.

\bibitem{Desh2} N. G. Deshpande, G. Eilam, X.-G. He 
and J. Trampetic, \Journal{\PLB}{366}{300}{1996}.

\bibitem{ciuchini} M. Ciuchini, E. Gabrielli, and G.F. Guidice,
CERN-TH-96-073, hep-ph/9604438 and private communication.
%
%
\bibitem{phix}
K. W. Edwards et al. (CLEO Collaboration),  CLEO CONF 95-8. 
%

\end{thebibliography}
\end{document}